# Zeroshot Listwise Learning to Rank Algorithm for Recommendation

HAO, WANG*

CEO Office, Ratidar Technologies LLC

Learning to rank is a rare technology compared with other techniques such as deep neural networks. The number of experts in the field is roughly 1/6 of the number of professionals in deep learning. Being an effective ranking methodology, learning to rank has been widely used in the field of information retrieval. However, in recent years, learning to rank as a recommendation approach has been on decline. In this paper, we take full advantage of order statistic approximation and power law distribution to design a zeroshot listwise learning to rank algorithm for recommendation. We prove in the experiment section that our approach is both accurate and fair.

## 1 INTRODUCTION

Recommendation is a must-have product for big corporations online. The click-through-rate of recommender systems is usually between 1% and 10%. For a website with more than 1 million page view per month, recommender systems can increase the page view by 30% - 40% (Amazon / Toutiao). If the website uses Google Ads to enlist users, it would cost much more money than building a recommendation team to increase the traffic volume.

There have been a large volume of literature on recommender systems since its debut in 1992. Prior to around 2015, the absolute majority of the algorithms are shallow models including collaborative filtering, matrix factorization, linear models, learning to rank and hybrid models. Unlike deep learning models in the following years, these models are much less complex in theory and implementation. The studies of these models are still on-going even today since they serve as simple and easy benchmarks for research.

---

* Place the footnote text for the author (if applicable) here.

Nearly every recommender system faces cold-start problem - how do we recommend items to users based on no input data ? This question sounds counter-intuitive to us since humans can not guess preferences of others if we do not know each other. However, this is a challenge that scientists and engineers must confront themselves with every day. A widely adopted technology in modern times is transfer learning / meta learning. However, since the debut of ZeroMat in 2021, a series of authentic zeroshot learning algorithms such as DotMat, RankMat, PoissonMat and PowerMat have emerged as a new school of algorithms that, for the first time in the history of AI, use no input data at all and solve the cold-start problem.

In this paper, we follow this line and propose a new learning to rank algorithm that falls in the category of listwise learning to rank paradigm. In 2023, a pairwise learning to rank algorithm named Skellam Rank was proposed. The algorithm uses input data structure, but can be transformed to a zeroshot learning algorithm very easily. The algorithm illustrated in this paper is the 2nd learning to rank algorithm solving the cold-start problem without reference to input data.

Our algorithm uses the theory of order statistics, power law theory and learning to rank design principles. We demonstrate the superiority of our approach in the Experiment section and prove the effectiveness of our design for the readers.

## 2 RELATED WORK

With the surge of internet applications such at Toutiao and TikTok, recommendation emerged as a symbolic technology in the 2010's. Deep neural models such as Wide & Deep [1], DeepFM [2] and DLRM [3] are widely deployed in online production systems. Although these methods seem effective, in recent years, scientist start to utilize large models [4][5][6] for improvement over older systems.

Recommender systems has many different research aspects. Fairness is one such topic that has attracted a lot of scientists in recent years. H. Wang [7] proposed to use tail index estimators to evaluate algorithmic performance on popularity bias effect. A. Beutel [8] came up with a regularization method to reduce the popularity bias problem. Other researchers take advantage of a whole spectrum of different approaches [9][10][11] to tackle the fairness problem.

Scenario-based recommendation is also a field that is in spotlight in recent years. Context-aware recommender systems (CARS) [12][13] can be deployed in multitudes of application scenarios ranging from fast food services to in-car music. Sequential recommendation [14][15] is another field that received best paper awards at research conferences such as RecSys.

Zeroshot learning problem is a challenge that scientists and researchers need to face on a daily basis. Conventional approaches include transfer learning [16][17] and meta learning [18][19]. In 2021 and 2022, a new school of techniques including ZeroMat [20], DotMat [21], RankMat [22] and PoissonMat [23] were invented so data-agnostic models emerged as a strong technique to tackle the problem.

## 3 ORDER STATISTICS

We find the following theorem [24] related to order statistics of utmost importance to our research:

**Theorem 3.1** Assume random variables $x_1, x_2, \cdots, x_n$ are i.i.d. distributed with density function f(x), the joint distribution of order statistics for a dataset containing n samples is the following :

$$f_{1,2,\cdots,n:n}(x_1, x_2, \cdots, x_n) = n! \prod_{i=1}^{n} f(x_i)$$

This theorem is the building stone of our listwise learning to rank algorithm in this paper.

## 4 LEARNING TO RANK

Learning to rank is a highly effective technique for ranking items for recommendation. Bayesian Personalized Ranking is one of the most notable learning to rank algorithm in the 2010's. The basic idea behind the approach is to maximize the following maximum likelihood estimator :

$$L = \prod_{i=1}^{N}\prod_{j=1}^{M}\prod_{k=1}^{N}\prod_{t=1}^{M} P(R_{i,j} > R_{k,t}) I(R_{i,j} > R_{k,t})$$

BPR-MF is a special case of Bayesian Personalized Ranking that uses matrix factorization to approximate the user item rating values R. MF is the shorthand for matrix factorization, whose loss function is the following : $L^{'} = \left(R_{i,j} - U_i^T \cdot V_j\right)^2$

The full loss function of BPR-MF is below:

$$L = \prod_{i=1}^{N}\prod_{j=1}^{M}\prod_{k=1}^{N}\prod_{t=1}^{M} P(U_i^T \cdot V_j > U_k^T \cdot V_t) I(R_{i,j} > R_{k,t})$$

The art of algorithmic design of this paradigm then becomes reduced to the formulation of $P(U_i^T \cdot V_j > U_k^T \cdot V_t)$. The classic paper of Bayesian Personalized Ranking uses logistic function to compute the probability, while Skellam Rank uses Skellam Distribution. The classic formulation of BPR-MF is not zeroshot learning, but Skellam Rank could be easily transformed into zeroshot learning.

In our formulation problem, we observe that the user item rating data in the input data structures follow power law distribution, so that we could use the values of rating data themselves to approximate the distribution of the values. For example, we use 5 to approximate the frequency of 5-star items, 4 to approximate the frequence of 4-star items, ..., etc. Therefore we can approximate the order statistics of the data distribution in the following way:

$$f_{1,2,\cdots,n:n;1,2,\cdots,m:m}(R_{1,1}, R_{1,2}, \cdots, R_{1,m}; \cdots; R_{n,1}, R_{n,2}, \cdots, R_{n,m}) \propto \prod_{i=1}^{n} \prod_{j=1}^{m} (U_i^T \cdot V_j)^{U_i^T \cdot V_j}$$

Using Stochastic Gradient Descent (SGD) to compute the optimal U and V for this estimator, we obtain the following results:

$$\frac{\partial L}{\partial U_i} = (U_i^T \cdot V_j)(U_i^T \cdot V_j)^{(U_i^T \cdot V_j - 1)} V_j + (U_i^T \cdot V_j)^{U_i^T \cdot V_j} log(U_i^T \cdot V_j) V_j$$

$$\frac{\partial L}{\partial V_j} = (V_j^T \cdot U_i)(V_j^T \cdot U_i)^{(V_j^T \cdot U_i - 1)} U_i + (V_j^T \cdot U_i)^{V_j^T \cdot U_i} log(V_j^T \cdot U_i) U_i$$

By carefully examining the formulas, we notice that there is no input data R in the formulas. The only variables in computation are the parameters U and V. Therefore the algorithm is a zeroshot algorithm.

We use the power function because the frequency of a user item rating value in the input data set is proportional to the rating value itself. We fully utilize the power law effect to design the listwise learning to rank algorithm.

## 5  EXPERIMENTS

We test our algorithm on MovieLens 1 Million Dataset and LDOS-CoMoDa Dataset. The accuracy metric we use in our experiments is MAE (Mean Absolute Error), and the fairness metric we use is Degree of Matthew Effect. We compare the zeroshot listwise ranking algorithm against ZeroMat / ZeroMat Hybrid, DotMat / DotMat Hybrid, classic matrix factorization and a couple of heuristics.

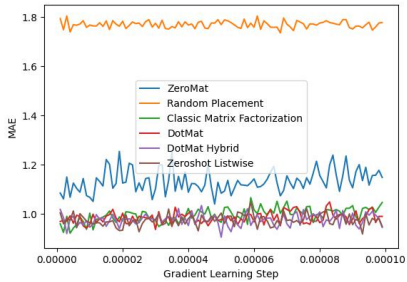

Figure 1 MAE comparison on MovieLens Dataset

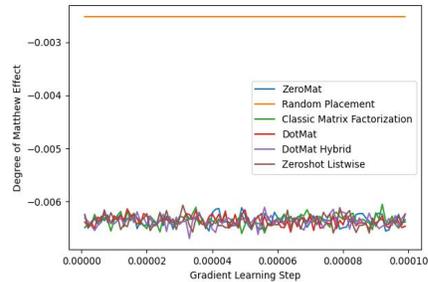

Figure 2 Fairness comparison on MovieLens Dataset

Figure 1 and Figure 2 demonstrate the effectiveness of the algorithm : On the MAE score evaluation test, the algorithm is the best one when the learning rate of SGD is selected properly. As for the fariness test, the algorithm is also highly competitive.

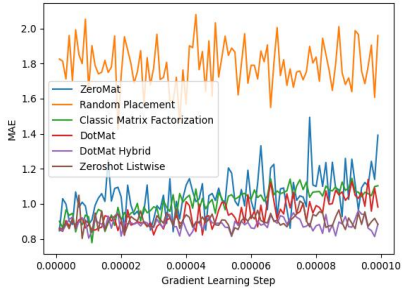
Figure 3 MAE comparison on LDOS-CoMoDa Dataset

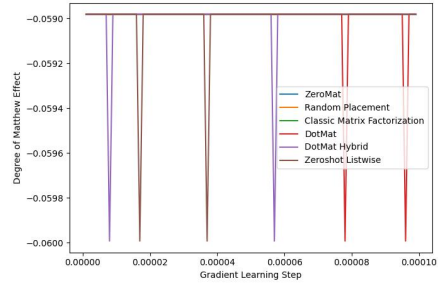
Figure 4 Fairness comparison on LDOS-CoMoDa Dataset

Figure 3 and Figure 4 demonstrate that the algorithm is competitive with DotMat Hybrid - an algorithm that takes advantage of full input data structures. The algorithm is also superior to other algorithms used in experiments.

## 6 CONCLUSION

In this paper, we propose a new zeroshot listwise learning to rank algorithm that takes no input data and generates highly effective results compared with other modern algorithms on both accuracy metric and fairness metric. In future work, we would like to explore the social science aspect of the algorithm so we could not only improve the internet product efficiency but also help our society advance.

# Authors' Information

| Your Name | Title* | Research Field | Personal website |
|---|---|---|---|
| Hao Wang | Director / CEO / Founder | Machine Learning / Computer Graphics | |
| Second Author | | | |
| Third Author | | | |